\documentclass[twocolumn,preprintnumbers,amsmath,amsfonts,amssymb,notitlepage,showpacs,aps,prb,longbibliography]{revtex4-2}

\usepackage{color}

\usepackage{amsmath}

\usepackage{graphicx}

\usepackage{hyperref}

\def \be{\begin{equation}}
\def \ee{\end{equation}}
\def \ba{\begin{array}}
\def \ea{\end{array}}
\def \bea{\begin{eqnarray}}
\def \eea{\end{eqnarray}}

\date{\today}
\begin{document}
\title{
Multipartite entanglement structures in quantum stabilizer states}

\author{Vaibhav Sharma}
\email{vaibhavsharma@rice.edu}
\thanks{Current Address: Department of Physics, Rice University, Houston, Texas}
\author{Erich J Mueller}
\email{em256@cornell.edu}
\affiliation{Laboratory of Atomic and Solid State Physics, Cornell University, Ithaca, New York}

%\affiliation{Laboratory of Atomic and Solid State Physics, Cornell University, Ithaca, New York}

\begin{abstract}

We develop a 
%new 
method for visualizing the internal structure of multipartite entanglement in pure stabilizer %quantum 
states. %Our scheme 
Our algorithm graphically organizes the many-body correlations in a hierarchical structure.  This provides a rich taxonomy from which one can simultaneously extract many quantitative features of a state including some traditional quantities such as entanglement depth, $k$-uniformity and entanglement entropy. Our method also presents an alternative computational tool for extracting the exact entanglement depth and all separable partitions of a stabilizer state.
%
%
%scheme includes prior approaches to characterizing multipartite entanglement, such as entanglement depth, while providing a rich taxonomy.
%organize the many-body correlations present in a state into correlation structure diagrams through a novel bottom-up approach. 
%In contrast to existing multipartite entanglement measures, these diagrams simultaneously provide the entanglement depth of the state (denoting what parts of the system are genuinely entangled) and an upper bound on the bipartite entanglement entropy across a partition. 
Our construction is
%Being 
gauge invariant and goes beyond traditional entanglement measures by visually revealing how quantum information and entanglement is distributed.
%in a multipartite entangled state.
%In particular, it the correlation structures reveal the complexity of stabilizer states by finding the length of many-body operators that encode information. 
We use this tool to analyze
%first apply this approach 
%to learn about 
the internal structures of prototypical stabilizer states (GHZ state, cluster state, stabilizer error correction codes) and are able to contrast the complexity of highly entangled volume law states generated by random unitary operators and random projective measurements.   
\end{abstract}

\maketitle

\section{Introduction}\label{intro}

Entanglement is %one of 
the key feature that distinguishes quantum and classical systems. It is a valuable resource 
%which forms the backbone of %several 
for
quantum information processing and computation~\cite{entanglement}. When restricted to two parties, entanglement is well understood: Bipartite entanglement measures are very intuitive, and there is no controversy about how to detect or quantify two-particle entanglement.
%in detecting and quantifiying the degree of entanglement. However f
For many-body systems, the situation is different.  Characterizing and understanding %true 
multipartite entanglement 
is
%remains 
challenging due to its far richer structure 
that cannot be easily summarized%
%and the near infinite possibilities of its characterization
~\cite{MA2024,dur,szalay,Walter2016,Barnum2001,miyake,coffman,goldbart}. 
Here we develop a more intuitive and broadly applicable approach towards organizing and visualizing multipartite entanglement in stabilizer states.

There exist several measures of multipartite entanglement such as the entanglement entropy~\cite{entanglemententropy}, entanglement depth~\cite{depth1,depth2}, global entanglement~\cite{global}, quantum fisher information~\cite{qfi}, Schmidt measure~\cite{Schmidt}, generalized geometric measure~\cite{ggm} and $N$-tangle~\cite{ntangle}. Each of these measures emphasize a different  aspect of multipartite entanglement, 
quantifying it with a single number. For example, the bipartite entanglement entropy across a given partition  quantifies the amount of entanglement between two regions. The entanglement depth specifies the number of entangled qubits in a given state, independent of the degree of entanglement. Quantum fisher information provides a lower bound on entanglement depth~\cite{qfi} while the Schmidt measure distinguishes between a product state and an entangled state. The $N$-tangle is able to numerically differentiate between some classes of many-body states like the GHZ state and the W-state~\cite{ntangle1}. A comprehensive explanation of these multipartite entanglement measures, as well as several others, is given in Ref.~\cite{MA2024}.  

Although these measures provide useful information, 
they fail to fully capture the complexity of a multipartite entangled state, particularly the internal structure of correlations and the local distribution of entanglement. This is especially apparent for states that are not described by traditional order parameters, such as those generated by noisy quantum circuits.   
%Our goal is to produce a hierarchical classification scheme which gives a 
%
%
%
%identify multipartite entanglement with a monotonic numerical quantity. However by characterizing a state with just a number, these measures are partial in their own way and often miss some key aspects of truly capturing multiparticle entanglement. 
%For example, bipartite entanglement entropy counts how many degrees of freedom in one part of the system are entangled with the rest of the system.  Entanglement depth quantifies the ways in which the system can be decomposed into unentangled subsystems.
%
%corresponds to the size of the largest subset of the system which is unentangled with the rest of the system, but does not contain any unentangled 
%
%
%only tells you how much information is shared between two parts of the system.
%doesn't indicate how many qubits in a given many-body system are genuinely multipartite entangled. 
%Entanglement depth specifies the number of qubits that are necessarily multipartite entangled in a system but fails to capture the amount/strength of entanglement. The Schmidt measure cannot distinguish between bipartite and multipartite entangled states. In the absence of an order parameter, these entanglement measures often don't give any insight into the internal properties of the state. 
In this work, our goal is to resolve the structure of an entangled state and develop a  way to visualize multipartite entanglement, going beyond reducing it to a single %monotonic 
numerical quantity. We propose 
%looking at 
organizing
multipartite entanglement via the topological structure of correlations present in the state. We group the qubits in a quantum state in clusters characterized by a parameter, $w$. The defining features of these $w$-clusters is that each qubit inside them has non-zero correlations with a subset of least $(w-1)$ other qubits within the same cluster. The idea is that $w$ qubits together encode information that is not present in any individual qubit.  We then recursively continue to group these clusters together, forming an entanglement structure diagram that shows how various parts of the system are connected. 
%This enables us to determine a state's complexity and gain better insights into the properties of the state. 
There is some relationship between this concept and the idea of the {\em information lattice}, which additionally incorporates spatial information \cite{artiaco2024universalcharacterizationquantummanybody,Klein,PRXQuantum.5.020352}.

%Using a state's many-body correlation functions, we 
%We hierarchically divide the qubits in a quantum state into clusters which are characterized by the lowest weight 

%The richness of multi-particle entanglement is already seen in the case of 3 qubits.

Our work has multiple motivations, both fundamental and practical.  For many years, various metrics have been invented to answer the question, “How do we define multiparticle entanglement?”  Our starting point is that there is no single metric of multiparticle entanglement that captures everything. Thus, we develop a scheme that can highlight its multifaceted nature. Our diagrams provide one such organizing structure that enables us to simultaneously extract different quantitative aspects of a multipartite entangled state. Following this approach, we are able to directly visualize how quantum information is spread throughout the system in  highly entangled states. We get access to the internal structure of correlations that connect various qubits in a composite many-body state.

We find that for many classes of states our entanglement diagrams can be efficiently calculated, though in the very worst case scenario our construction algorithm requires order $L\choose{L/2}$ operations on a system of size $L$.   Even in this worst case, our construction may be useful. Our diagrams provide bounds on the bipartite entanglement entropy across {\em all} possible partitions of the system, using significantly fewer resources than directly calculating each of those entropies.
%We note that the recursive nature of our method makes extracting such information much more efficient than simply calculating bipartite entanglement entropy across all possible partitions of a given system. 
Our method also %helps us extract entanglement measures like 
provides the
entanglement depth, $k$-uniformity, and a number of other metrics.
%and an upper bound to entanglement entropy across any partition. 

%which are the class of quantum states which are the simultaneous eigenstates of . Stabilizer states have an efficient classical representation and all correlations are binary valued, represented by strings of Pauli operators. The correlation structures are gauge-invariant quantities and hence reflect the complexity of a state by revealing the true length of many-body operators that encode information. We focus on stabilizer states of qubits arranged in a 1D chain. The correlation diagrams reveal interesting spatial structures of some well-known stabilizer states like the cluster state, GHZ state and error correction codes such as the 5-qubit code and the 7-qubit CSS code. 

We illustrate our decomposition by looking at well-known stabilizer states:  the cluster state, GHZ state and the logical states of error correction codes such as the five-qubit code and the seven-qubit CSS (Calderbank-Shor-Steane) code.  The true power of our construction, however, is revealed by looking at 
%We also use the diagrams to analyze 
highly entangled volume law states generated by random two-qubit Clifford unitaries and random three-qubit projective measurements \cite{trans2,measonly1}. Although both sets of states show volume law scaling of bipartite entanglement entropy,
%on a 1D chain of qubits. W
we find
that the entanglement structure diagrams are able to distinguish
between them. 
%We explain the algorithm to extract the entanglement structure diagrams. Finally we discuss how the computation time to make the diagrams scales with system size, particularly for highly entangled states generated by unitaries and measurements. 

\section{Algebraic Structure}\label{sec:algebraicstructure}

Our approach works for stabilizer
states \cite{gottesman}, which are 
%an important subclass of quantum states.  %best suited for pure 
%They are 
an important class of quantum states that are used
in quantum computing and play an important role in quantum error correction.
An $N$ qubit stabilizer state is the simultaneous eigenstate of $N$ %linearly 
independent commuting Pauli strings, referred to as the stabilizer generators.  For example, the cat state, $1/\sqrt{2}(|111\rangle+|000\rangle)$ is the simultaneous eigenstate of three Pauli string operators, $Z_1Z_2I_3$, $I_1Z_2Z_3$ and $X_1X_2X_3$.  Here $X_j,Y_j,Z_j$ are the Pauli operators $\sigma_x^j,\sigma_y^j,\sigma_z^j$ acting on qubit $j$, $I_j$ is the identity matrix, and a Pauli string is a product of one of these four operators acting on every qubit.  Trivially, a stabilizer state is also an eigenstate of any member of the {\em stabilizer group} formed by taking arbitrary products of the generators.  The stabilizer generators are not unique, and can be replaced by any $N$ independent elements of the stabilizer group.  This freedom to choose the generators is often referred to as a {\em gauge} freedom.   We %will 
only consider {\em pure} quantum states, and will not discuss the properties of ensembles encoded by density matrices.

A generic Pauli string can be written as 
$P=\prod_j X_j^{\alpha_j} Z_j^{\beta_j}$, where $\alpha_j,\beta_j=0,1$.
The {\em support} of a Pauli string is the set of qubits for which  $\alpha$ and/or $\beta$ is non-zero.  The {\em weight} of the string is the number of qubits in the support.
The expectation value of a weight-$w$ stabilizer operator can be interpreted as the expectation value of a $w$-spin correlation function.  
In our approach, we 
%will be 
recursively group sets of qubits into clusters.  At each iteration, we %will 
%treat 
take
the weight of a stabilizer operator to be 
%given by 
the number of clusters in its support, rather than the number of qubits.

 A key feature of stabilizer states is that the bipartite entanglement entropy is {\em quantized}.  If we break a stabilizer state into two disjoint sets of qubits, the entanglement entropy across the cut is always a multiple of $\ln(2)$.  If region $A$ contains $n_A$ qubits then $S_A\leq n_A\ln(2)$.  When this bound is saturated, we refer to $A$ as being {\em maximally entangled}.  One learns nothing about the state of the system by interrogating region $A$.  There is no information which is exclusively stored in $A$, and the reduced density matrix is the identity matrix of dimension $2^{n_A}$. 
 
 A useful intuition is that if $S_A=s_a\ln(2)$ with respect to the rest of the system, then there are %effectively 
 $n_A-s_a$ bits of information that are locally stored within $A$.  In particular, it is an eigenstate of exactly $n_A-s_a$ %linearly 
 independent Pauli strings whose support is entirely within $A$.  These are the operators which measure the stored information.
 %in $A$.  
 They generate a subgroup of the stabilizer group, whose elements we refer to as the $A$-stabilizer operators.  Another useful feature is that the reduced density matrix of $A$ has rank $2^{s_a}$.

We can generalize this notion to the case when region $A$ is divided into disjoint clusters of qubits, constructed by some as-yet unspecified algorithm.  Each cluster will be labeled by an index $i$.  Following Watanabe \cite{Watanabe}, we define the  {\em total correlations} in $A$
to be 
$I_{A} = \left(\sum_{i\in A} S_i\right) - S_A$
where $S_i$ is the entanglement entropy of cluster $i$ with the rest of the system.  
%In the special case where each cluster contains a single qubit, $I_{A}=n_A-S_A$.  
The total correlations are also referred to as the ``multipartite quantum mutual information" \cite{Kumar}.  It tells us how many bits of information are stored in $A$, but which are not stored solely by any single cluster within $A$.

%with the remainder of the system.  The total correlation carries information about 

%a multipartite information quantity for the region $A$ called the {\em total correlation}. If all the qubits in region $A$ have entanglement entropy $S_i$ with the rest of the system, the total correlation is given by, $I_{A} = \sum_i S_i - S_A$. The interpretation is similar to the case where the entanglement entropy bound for region $A$ is not saturated.  When $ I_{A} \neq 0$, there are some bits of information stored when we interrogate all the qubits in $A$ denoting stabilizers with support only within region $A$. 

% Equivalently, the subgroup of the stabilizer group whose support is in $A$ is generated by $n_a-s_a$ elements. 

In this work, we use the term: {\em $w$-cluster}. We say that region $A$ is a {\em $w$-cluster} if it fulfills the following two conditions: (1) Any subset  $C\subset A$ containing $w-1$ or fewer elements has vanishing total correlations, $I_C=0$. Equivalently, any stabilizer operators whose support lies solely in $A$ must have a
weight greater or equal to $w$. (2) For every element $j\in A$, there exists a set $B\subseteq A$ containing $w$ elements such that $B$ contains $j$ and has non-zero total correlations. Furthermore, 
$B$ cannot be disjoint from all other such $w$-element subsets within $A$.  Equivalently,  every element of $A$ is in the support of an $A$-stabilizer operator of weight less than or equal to $w$.

%three new terms: {\em $w$-indivisible}, {\em $w$-connected} and {\em $w$-cluster}. 
%We say that region $A$ is {\em $w$-indivisible} if any subset  $C\subset A$ containing $w-1$ or fewer elements have vanishing total correlations, $I_C=0$.  Equivalently, any stabilizer operators whose support lies solely in $A$ must have a
%$A$-stabilizers must have 
%weight greater or equal to $w$. 

%We say that $A$ is {\em $w$-connected} if  every element $j\in A$ belongs to a set $B\subseteq A$ containing $w$ elements, such that $B$ has non-zero total correlations.  Furthermore, 
%$B$ cannot be disjoint from all other such $w$-element subsets within $A$.  Equivalently,  every element of $A$ is in the support of an $A$-stabilizer of weight less than or equal to $w$.
%The region $A$ is a {\em $w$-cluster} if it is both {\em $w$-indivisible} and {\em $w$-connected}.

In our construction, we use an iterative method to divide the system into $w$-clusters. In the first iteration, each qubit is considered as a single element. At each subsequent iteration, any $w$-clusters found in the previous iteration are treated as single indivisible elements. For every iteration, we begin by finding all of the elements which are disentangled from the rest of the system.  These are 1-clusters that are decoupled from the rest of the system and can be ignored at future steps.  Next we find all of the elements which belong to 2-clusters.
At this point we treat the clusters as indivisible elements. If any 2-clusters are found, we go to the next iteration and repeat the previous steps -- removing single elements which are decoupled, and forming new 2-clusters of the given elements. During any iteration, if no new 2-clusters can be formed and there are still elements which have not yet been assigned to decoupled clusters, we search for 3-clusters.  If any 3-clusters are found, we go to the next iteration and we repeat all the previous steps starting from finding decoupled singletons and new 2-clusters. If no 3-clusters are found, we move onto finding 4-clusters, 5-clusters and so on until we find some cluster with non-zero total correlations. We always construct as many low weight clusters as possible before moving on to higher weight clusters. We continue iterating until all qubits have been assigned to large decoupled clusters. We 
give more details of the
%describe the 
computational procedure %steps in detail 
in Appendix~\ref{algorithm}.

\begin{figure}
\includegraphics{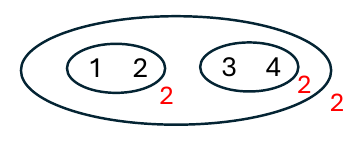}
    \caption{Entanglement structure diagram of a four qubit state given by the wavefunction, $|\psi\rangle = 1/2(|1111\rangle +|0011\rangle+|1100\rangle-|0000\rangle$). The qubits, labeled by black integers, are placed in clusters denoted by drawing circles around them. The clusters are then considered indivisible entities and can be further placed into bigger clusters. The small red number, $w$ in the bottom left of a clustering circle denotes that the entities inside that cluster share a minimum of $w$-point correlation function among the entities. }
    \label{fig:fourqubit}
\end{figure}

A short example is useful.  Consider the four-qubit state, $|\psi\rangle$ with stabilizer generators $Z_1 Z_2$, $Z_3 Z_4$, $X_1 X_2 Z_3$ and $Z_2X_3X_4$.  Up to a normalization constant, $|\psi\rangle=|1111\rangle +|0011\rangle+|1100\rangle-|0000\rangle$. The entanglement structure of this state is shown in Fig.~\ref{fig:fourqubit}. Here the qubits are labeled by black integers. The $w$-clusters are denoted by ovals, with $w$ shown in red. In the first round, our algorithm produces two 2-clusters: $(1,2)$ and $(3,4)$.  In the second round, we consider the clusters $(1,2)$ and $(3,4)$ as single indivisible elements.  We then see that these together form one large 2-cluster: $[(1,2),(3,4)]$ connecting all four qubits into a single many-body state. Such a structure implies that while qubits $1,2$ and $3,4$ are directly connected by weight-2 stabilizer operators, qubits $1,3$ and qubits $1,4$  are only connected by higher weight stabilizer operators. 
%They are further joined when they are part of the clusters $(1,2)$ and $(3,4)$. 
In this way our construction recursively forms clusters to build a complete description of  entanglement within the many-body state.

\begin{figure}
\includegraphics[width = \columnwidth]{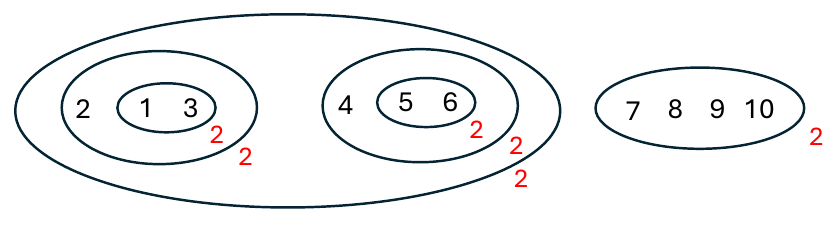}
    \caption{An entanglement structure diagram of an arbitrary ten-qubit stabilizer state is shown. Its stabilizer generators are written in Appendix~\ref{app:stabilizers}. The state is separable with qubits labeled 1-6 forming the largest cluster, corresponding to an entanglement depth of 6. The nested structure of the clusters shows how correlations are distributed in the state.}
    \label{fig:struct}
\end{figure}

Figure~\ref{fig:struct} shows a more complicated entanglement structure diagram of a ten-qubit state. The corresponding stabilizer group is written in Appendix~\ref{app:stabilizers}. In this diagram, qubits labeled 7, 8, 9 and 10 form an independent cluster which is unentangled with the rest of the system. It is labeled as a 2-cluster, indicating that each of these qubits must be in the support of a weight 2 stabilizer operator, whose support is fully within this cluster.

The cluster containing qubits from 1-6 is the largest, with multiple sub-clusters. At the first level, qubits (1,3) and qubits (5,6) are placed in 2-clusters, denoting the presence of weight 2 stabilizer operators.  By virtue of being 2-clusters that are not isolated, each of them contains one bit of information and has an entanglement entropy $S=\ln 2$ with the rest of the system.

In the next iteration, the 2-cluster formed by qubits (1,3) and qubit 2 join together to form a bigger 2-cluster. We can further deduce constraints from the fact that (1,3) is a 2-cluster embedded in a larger 2-cluster. This implies that there is a length 3 stabilizer operator whose support lies on $(1,2,3)$. If there were only length 2 stabilizer operators, $(1,2,3)$ would have all been placed in a 2-cluster in the first iteration.
Similarly, the 2-cluster of qubits (5,6) joins with qubit 4 to form a bigger 2-cluster.
 
Finally, the clusters of qubits [(1,3),2] and qubits [4,(5,6)] further join to form an independent 2-cluster containing qubits 1-6. The final clustering denotes a stabilizer operator whose support is in both of these clusters but that cannot be individually confined to either one of these.

\subsection{Entanglement depth}
Having constructed an entanglement diagram, we can readily extract the {\em entanglement depth} \cite{depth1,depth2}.
Any $n$-qubit pure state can be written as a product state over $m$ disjoint regions in the following form: 
\begin{equation}
    |\psi\rangle = |\phi_1\rangle_{A_1} \otimes |\phi_2\rangle_{A_2}..... \otimes |\phi_m\rangle_{A_m}
\end{equation}
where none of the $|\phi_j\rangle_{A_j}$ can be further decomposed into product states. The sets $A_1,A_2,\cdot A_m$ are the {\em separable partitions}.
If there are at most $k$ qubits in any of the $m$ partitions, the state is deemed to have an entanglement depth of $k$~\cite{depth1,depth2}. The  state is said to have $k$-qubit
 multiparticle entanglement. 
In our entanglement structure diagram, the entanglement depth is simply the number of qubits in the largest cluster. Our diagram also directly provides all the {\em separable partitions} of the state.

In the four qubit example shown in Fig.~\ref{fig:fourqubit}, all the qubits are ultimately part of one big cluster, denoting an entanglement depth of 4. In contrast, the diagram in Fig.~\ref{fig:struct} shows that the ten-qubit state can be written as a product state of the form, $|\psi\rangle = |\phi_1\rangle_{A_1} \otimes |\phi_2\rangle_{A_2}$ where region $A_1$ contains qubits labeled from 1 to 6 and $A_2$ contains qubits labeled from 7 to 10. Clearly this state has an entanglement depth of 6. 

Our construction provides an alternative way of calculating the exact entanglement depth as well as providing all the separable partitions of a given state. This approach is markedly different from entanglement witness based approaches such as quantum fisher information~\cite{qfi}, quantum squeezing~\cite{depth1} and optimized $n$-partite witnesses~\cite{witness} that only provide lower and upper bounds to entanglement depth.

Naively, one would need to do a large combinatorial search to find all separable partitions and extract the entanglement depth. This would always scale exponentially with system size. 
In Appendix~\ref{app:comptime}, we show that for many types of states it is computationally efficient to construct our entanglement structure diagrams.

\subsection{Minimal stabilizer weight and $k$-uniformity}\label{sec:minweight}

An important feature that can be extracted from these diagrams is the weight of the smallest stabilizer operator in the stabilizer group.  We refer to this as the {\em minimal stabilizer weight}. It reveals the degree of delocalization of  the quantum information.  In the entanglement structure diagram,  one can identify the minimal stabilizer weight as the smallest $w$, for the $w$-clusters in the first iteration of our algorithm. 

The minimal stabilizer weight has previously been discussed in the context of $k$-uniformity \cite{kuniform1,kuniform2,kuniform3,kuniform4,kuniform5}. Pure states of $N$ qubits are called $k$-uniform if the reduced density matrix of any subset of $k$ qubits or less is maximally mixed. A state with minimal stabilizer weight, $w$ must be $(w-1)$-uniform, and vice-versa. In a state with minimal stabilizer weight, $w$, all subsets of $(w-1)$ or fewer qubits have vanishing total correlations. All such subsets are thus maximally entangled with the rest of the system, making their reduced density matrices fully mixed. The minimal stabilizer weight is $2$ for both the states shown in Figs.~\ref{fig:fourqubit} and ~\ref{fig:struct}, making them $1$-uniform states.

\subsection{Bipartite entanglement entropy}\label{sec:eentropy}

We can also use the diagram to obtain an upper bound on the bipartite entanglement entropy across any given partition. Consider a set of $n$ qubits that are partitioned into two regions, region $A$ containing $m$ and region $B$ containing $n-m$ qubits with $m < n-m$. The maximum entanglement entropy across the partition in a stabilizer state is $m\ln 2$ since there are only $m$ units of information available in region $A$ that can be shared with the other region $B$. Within region $A$, any local clusters consisting of qubits entirely within the region reduce the entanglement entropy across the partition by at-least one unit. This is because such a local cluster denotes a stabilizer operator that is completely confined within region $A$. Identifying such local clusters gives upper bounds to the entanglement entropy of a given region with the rest of the system.

The four qubit example of Fig.~\ref{fig:fourqubit} is the easiest to understand. Consider region $A$ with qubits $(1,2)$ and region $B$ with qubits $(3,4)$. Simply by counting the number of qubits we know that the maximum possible bipartite entanglement entropy between $A$ and $B$ is $2\ln 2$. Beyond that, within region $A$ there is one local cluster containing one unit of information entirely within $A$. This reduces the entanglement entropy across the partition by $\ln 2$. Thus we know that the entanglement entropy across the partition is bounded from above by $\ln 2$.  In this case the bound is saturated, as the two subclusters belong to one bigger cluster, and hence share one bit of information.

%to be $S = 1*\ln 2$ which matches the entanglement entropy calculated using the stabilizer generators. .

Now consider the example shown in Fig.~\ref{fig:struct}, where qubits 1-6 are in a pure state.  We can separate that pure state into two regions -- for example, we take region $A$ to contain qubits 1-3 and region $B$ to contain qubits 4-6. From counting qubits, the maximum possible entanglement entropy across this partition is $3\ln 2$. Within region $A$, we see one sub-cluster that is a 2-cluster containing qubits (1,3). This reduces the entanglement entropy across the partition by $\ln 2$. It is further joined with qubit 2 in another local 2-cluster, reducing the entropy further by $\ln 2$. Thus the upper bound on entanglement entropy between regions $A$ and $B$ is $\ln 2$. 

We emphasize that the entanglement structure diagram is not a substitute to calculating the exact entanglement entropy across a given partition. That can be more efficiently done for stabilizer states by finding the binary rank of the reduced stabilizer tableau~\cite{trans3}. The utility of our construction is in providing a visualization of the upper bound to entanglement entropy across all partitions simultaneously.

\subsection{Entanglement structure layers}

Another quantity that can be easily extracted from the diagrams is the number of layers in the entanglement structure. In our recursive construction, we group qubits into clusters and then iterate to further group clusters together until all qubits have been assigned to large decoupled clusters.  The resulting diagrams thus have an onion-like structure with different layers corresponding to the different clustering iterations.
The example shown in Fig.~\ref{fig:fourqubit} has two layers. In the state shown in Fig.~\ref{fig:struct}, there are two large decoupled clusters. The cluster containing qubits labeled 1-6 has three layers. The cluster containing qubits 7-10 has only a single layer. The presence of multiple layers demonstrates a hierarchy in stabilizer weights and spatially local nature of correlations encoded by the stabilizer operators.

\section{Entanglement structures of some prototypical states}

\begin{figure}
\includegraphics[width = \columnwidth]{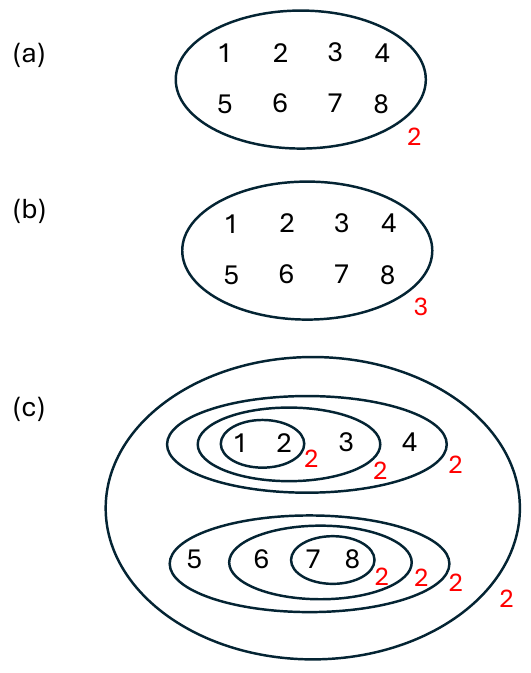}
    \caption{Entanglement structure diagram of 8-qubit (a) GHZ state, (b) 1D cluster state with periodic boundary conditions and (c) 1D cluster state with open boundary conditions}
    \label{ghzcluster}
\end{figure}

We can use the entanglement structure diagrams to visualize some prototypical stabilizer states. The diagram for a GHZ state \cite{ghz} of eight qubits, $|\psi\rangle=1/\sqrt{2}(|11111111\rangle+|00000000\rangle)$ is shown in Fig.~\ref{ghzcluster}(a). Among the eight stabilizer generators of the GHZ state, seven can be taken to have the form, $S_i = Z_i Z_{i+1}$, where $i \subset [1,.,7]$. The eighth generator can be chosen to be $S_8 = X_1 X_2 ...X_8$. As already introduced, $Z_i,X_i$ are Pauli operators $\sigma^z_i,\sigma^x_i$ for the qubit labeled $i$. All the qubits share one big 2-cluster, denoting the presence of 2-point correlations (weight-2 stabilizer operators) and a full entanglement depth of 8. In fact, in a GHZ state, each qubit pair (i,j) is in the support of a  weight-$2$ stabilizer operator, $Z_i Z_j$. The state is $1$-uniform with a minimal stabilizer weight of 2.

In Figs.~\ref{ghzcluster}(b), and ~\ref{ghzcluster}(c) we show the entanglement structure diagram for the eight-qubit cluster state on a one-dimensional (1D) lattice with and without periodic boundary conditions respectively \cite{cluster}. The cluster state is defined by its stabilizer generators:
$S_j = Z_{j-1}X_j Z_{j+1}$
for $j\neq 1,8$. For periodic boundary conditions, $S_1 = Z_8X_1 Z_2$, $S_8 = Z_7 X_8 Z_1$. The entanglement structure in this case contains all qubits in one big $3$-cluster due to all the weight-3 stabilizer operators. Without the periodic boundary conditions one instead has $S_1 = X_1 Z_2$, $S_8 = Z_7 X_8$. These boundary stabilizer operators have weight 2, and hence both $(1,2)$ and $(7,8)$ are 
placed in 2-clusters. 
Treating these 2-clusters as indivisible, the remaining stabilizer operators have the same topological structure. Thus $[(1,2),3]$ and $[6,(7,8)]$ form 2 clusters.  We can repeat, until one large cluster is formed. This largest cluster has four layers in the non-periodic boundary condition case.  As with the GHZ state, the entanglement depth is 8 in both the cluster states.

The presence of periodic boundary conditions significantly alters the entanglement structures of the cluster states. Most importantly, in the non-periodic case there exists some information which can be probed by interrogating only two qubits at the boundary. This is denoted by the presence of $2$-clusters at the boundaries. The state is $1$-uniform with minimal stabilizer weight of 2. In contrast, the periodic boundary case requires interrogation of at least three qubits to learn any information as all the qubits only come together in a $3$-cluster. The state is $2$-uniform with minimal stabilizer weight 3. Moreover the nested entanglement structure also better demonstrates the local nature of correlations as compared to the periodic boundary condtion case.

%Each of the two boundary clusters share a 2-point correlation with a neighboring site and start forming bigger 2-clusters until all qubits get absorbed into one big 2-cluster. The diagram shows the spatially local correlations of a cluster state. The state has a full entanglement depth of 8.

Both the GHZ and the  cluster state are low entanglement area law states. We know that across a bipartition in the center, they only have $\ln (2)$ entanglement entropy yet the entanglement diagrams have very different structures -- particularly in the absence of periodic boundaries. The $\ln(2)$  entanglement of the cluster state without periodic boundary conditions can be extracted from the reasoning in Sec.~\ref{sec:eentropy}.  We imagine a bipartition into the regions $A=(1,2,3,4)$ and $B=(5,6,7,8)$.  We can iteratively find the total correlations of region $A$:  There is one bit of information stored in $(1,2)$.  Every time we grow that cluster, we add one qubit, but also add one confined stabilizer operator.  Hence $[(1,2),3]$  and $[((1,2),3),4]$ each contain one bit of information.  Consequently the bipartition into $A$ and $B$ has an entanglement entropy of only $\ln(2)$.
The entanglement entropy of the GHZ state and the cluster state with periodic boundary conditions, however, cannot be extracted solely from the entanglement structure diagram.

%also contains one bit of information.
%'s diagram makes it easily apparent. At each iteration, all clusters are local and do not cross the center. Only at the final iteration, a 2-cluster joins qubits from each half, denoting a single unit of quantum information shared across the boundary. In the GHZ state, there are no multiple iterations since all qubit pairs are part of 2-clusters by being the support of a weight-2 stabilizer. The diagram by itself does not indicate that the GHZ state has $\ln 2$ entanglement entropy. Typically more iterations lead to a better upper bound on entanglement entropy. 

\begin{figure}
\includegraphics{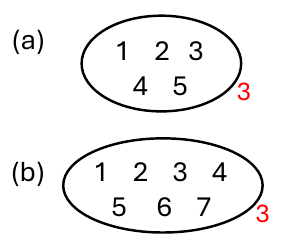}
    \caption{Entanglement structure diagram of (a) the five-qubit error correcting code and (b) seven-qubit Steane CSS error correcting code}
    \label{errorcorrect}
\end{figure}

Figures ~\ref{errorcorrect}(a) and ~\ref{errorcorrect}(b) show the entanglement structure diagrams of the five-qubit and seven-qubit error correcting codes. 
The stabilizer generators are given in Ref.~\cite{gottesman} and
these codes can correct arbitrary single qubit errors.
%~\cite{gottesman}. 
The figures show the diagrams in the case when the logical qubit is in the eigenstate of the logical $X$ operator. In both cases, all the qubits are in one 3-cluster showing an entanglement depth of 5 and 7 respectively. We can conclude that each of the qubits is in support of a weight 3 stabilizer operator that can be used to interrogate whether the system is in the code space. Both states are $2$-uniform with a minimal stabilizer weight of 3.  No extra information is given by the entanglement structure diagram, and one can argue that these diagrams have limited utility for such a class of states.

\section{Entanglement structures of highly entangled volume law states}\label{sec:volume}

We now analyze the internal structure of highly entangled volume law states generated in random quantum circuits. In volume law states, the entanglement entropy between regions of size $L$ scales as $L^d$ (volume of the region) while it scales as $L^{d-1}$ (area of the boundary) for area law states. Here $d$ is the spatial dimension. Random quantum circuits have recently been instrumental in enhancing our understanding of generic quantum dynamics and novel entanglement phase transitions between area and volume law phases~\cite{randomcircuits}. In 1D systems, a phase transition between area law and volume law phases occurs when single site projective measurements are interspersed with random two-qubit unitary operators~\cite{trans1,trans2,trans3,trans4}. 
%on a 1D chain of qubits  
Furthermore, there have been various measurement-only phase transitions from volume law to area law~\cite{measonly1} or between different area law phases using random non-commuting projective measurements~\cite{measonly2,measonly3,measonly4,measonly5,measonly6,measonly7,measonly8}. The states  generated by 
typical
random quantum circuits  
do not have an order parameter and are 
%often simply 
characterized by the scaling of their bipartite entanglement entropy. Not much is known about their spatial structure or properties. The entanglement structure diagrams can fill this gap.  

\begin{figure}
\includegraphics[width = \columnwidth]{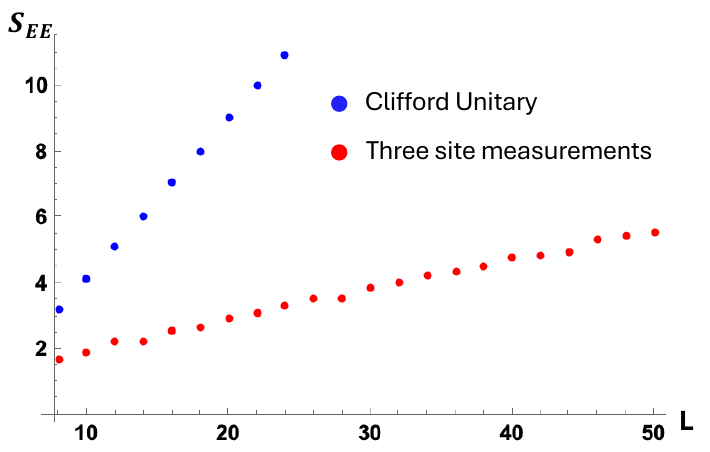}
    \caption{Average bipartite entanglement entropy, $S_{EE}$ as a function of system size $L$ for states generated by two-qubit random Clifford unitary operators (blue) and three-qubit random measurements (red). $S_{EE}$ scales linearly with $L$ in both cases.}
    \label{entropy}
\end{figure}

\begin{figure}
\includegraphics[width = \columnwidth]{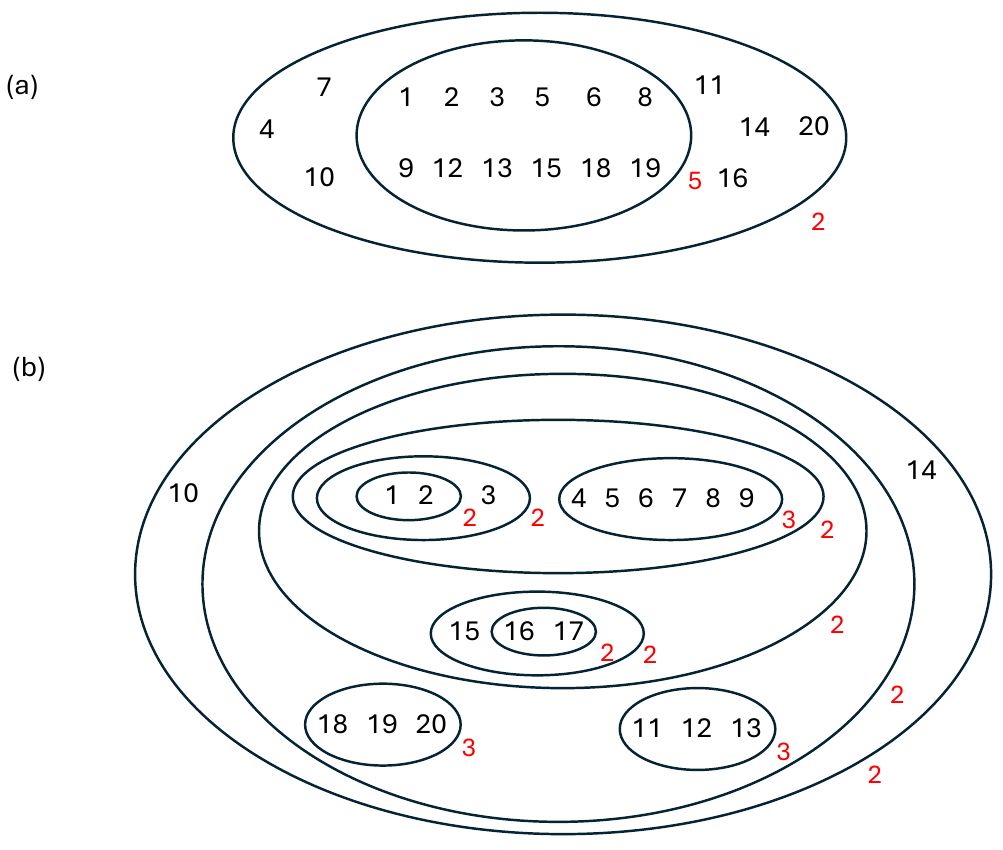}
    \caption{Entanglement structure diagram of a typical 20 qubit volume law state generated by (a) local two-qubit random Clifford unitary gates and (b) local three-qubit random projective measurements }
    \label{volumelawstruc}
\end{figure}

We focus on volume law states in one-dimension generated by random local two-qubit Clifford unitary gates or random local three-qubit projective measurements.  We take these measurements to be random weight-3 Pauli strings acting on neighboring sites. In our analysis, we start from a product state and apply our random unitary gates or measurements until a steady state distribution is produced.  We generate 500 states in both categories for various systems sizes, $L$. In Fig.~\ref{entropy}, we plot the average bipartite entanglement entropy, $S_{EE}$ between two halves of the system as a function of system size $L$. In both cases, $S_{EE} \propto L$, signaling a volume law. For the same system sizes, $S_{EE}$ is greater for the unitarily evolved state. In fact, the states generated by unitary evolution typically saturate $S_{EE}$ such that, $S_{EE} \sim L/2$. These are referred to as Page states \cite{page}.  This maximal entanglement entropy reflects the behavior of a quantum state under chaotic evolution where the quantum information is maximally scrambled. In contrast, the states generated purely by measurements have a lower entanglement entropy, but still follow the volume law. We will see how the entanglement structure diagrams can help us to visualize this difference.
%of entanglement entropy. 

\begin{figure}
\includegraphics[width = \columnwidth]{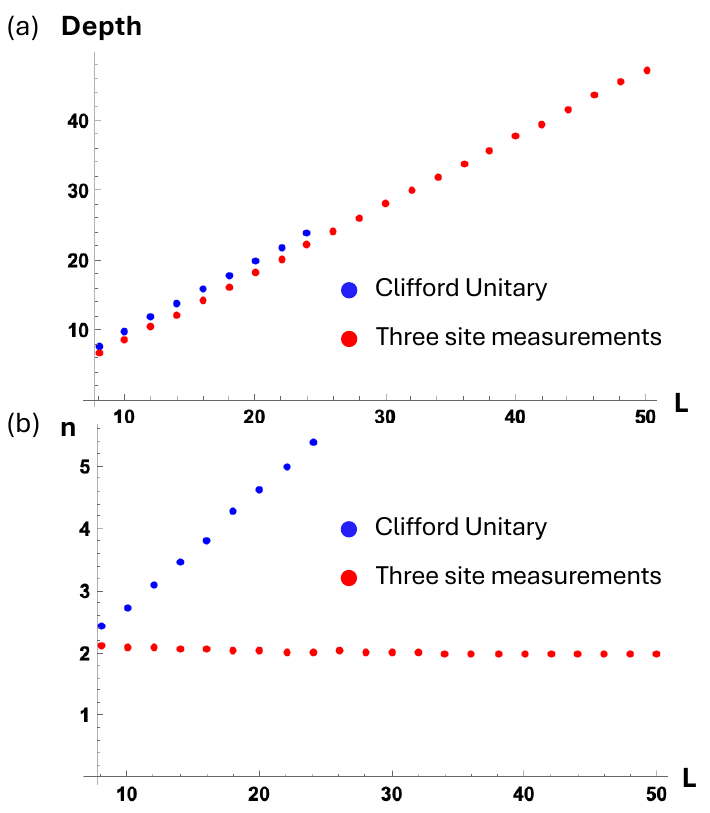}
    \caption{(a) Average entanglement depth and (b) minimal stabilizer weight $n$ of volume law states generated by two-qubit random Clifford unitary operators (blue) and three-qubit random measurements (red). Both kinds of volume law states have similar entanglement depth. However the minimal stabilizer weight found in the entanglement structure diagram scales linearly for unitarily evolved states while it remains constant for measurement-only states}
    \label{depthandn}
\end{figure}

We first generate some entanglement structure diagrams to learn about the internal structure of these volume-law states. Figure~\ref{volumelawstruc} shows the entanglement structure diagram of a typical unitary and measurement-only volume law state of 20 qubits. The measurement-only state has more layers with a lot of local sub-clusters that do not have correlations going across the boundary at the middle. These local sub-clusters dilute the long-range spreading of quantum information and reduce the bipartite entanglement entropy. The minimal stabilizer weight is $2$, implying that it is  a $1$-uniform state. On the other hand, the unitary volume law states have no such local sub-clusters, suggesting a more complete scrambling of information and a larger (maximal) entanglement entropy. For this example, the minimal stabilizer weight is $5$, making this state $4$-uniform.

The multifaceted nature of the entanglement structure diagrams lets us extract a variety of quantitative features from these states. For instance, we can look at entanglement depth, minimal stabilizer weight, the spatial range of the minimal weight stabilizer operators, number of layers in the diagram, average size of clusters, etc. Using some of these features, we would now show that the differences we see in the entanglement structure diagrams of these two typical states are not incidental. They are signatures of the statistical differences between the entire ensemble of such states.

We construct the entanglement structure diagrams for the 500 unitary and measurement-only volume law states that we generated earlier for various system sizes, $L$. We then extract ensemble averages of various quantities. Figure~\ref{depthandn}(a) shows the average entanglement depth and Fig.~\ref{depthandn}(b) shows the minimal stabilizer weight, $n$ [see Sec.~\ref{sec:minweight}].  Both categories of states show a full entanglement depth, $\sim L$, and this metric cannot be used to distinguish them.  They do, however,
differ sharply in their minimal stabilizer weight. 
This weight grows linearly with system size for volume law states generated by unitary operators while it stays constant for states generated by measurements only.

The minimal stabilizer weight can also be interpreted as the $k$-uniformity of the state. For unitary volume law states, $k$ grows linearly with system size. For measurement-only volume states, the states are typically $2$-uniform with $k \sim 2$ regardless of system size. The information stored in the unitary volume law states is more delocalized -- extracting any information requires measuring an extensive number of qubits.
%the correlation  have more scrambled information since the correlation functions become longer as the system size increases. 
As is detailed in Appendix~\ref{app:comptime}, the large minimal stabilizer weight makes it expensive to construct entanglement diagrams for the unitary volume law states.  Thus we are limited to moderate system sizes, 
%We can only create entanglement diagrams of moderate system sizes, 
$L \sim 24$ for unitary volume law states due to this increasing complexity. 

\begin{figure}
\includegraphics[width = \columnwidth]{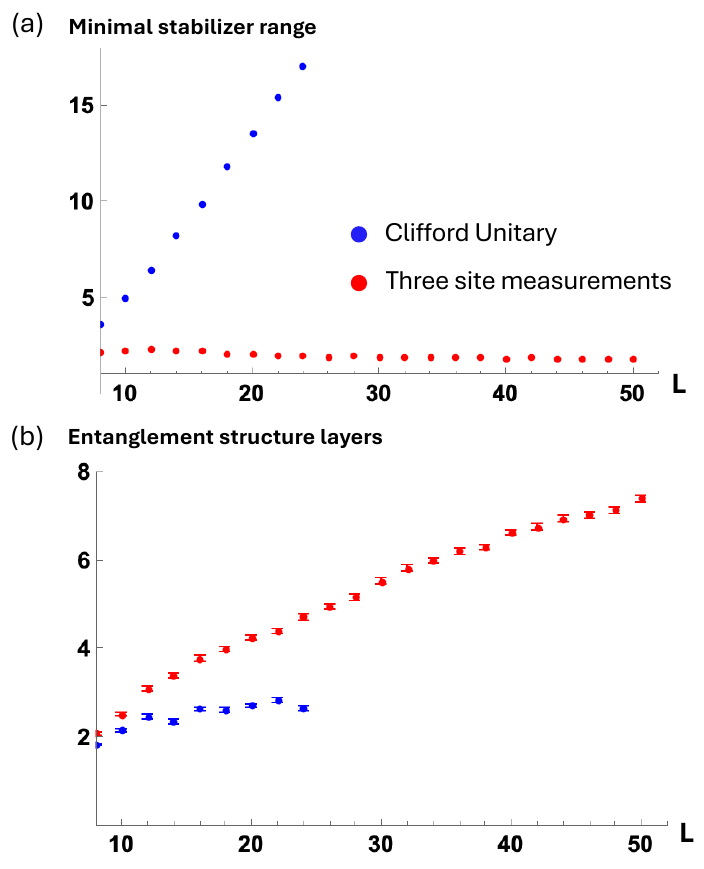}
    \caption{(a) Average spatial range of minimal weight stabilizer operators as a function of system size $L$. It grows linearly with system size in unitary volume law states (blue) while remaining nearly constant for measurement-only volume law states (red). (b) Average number of entanglement structure layers as a function of system size $L$. The number of layers stay nearly constant for unitary volume law states (blue) while they grow with system size for measurement-only volume law states.}
    \label{rangeandnumlayers}
\end{figure}

In Fig.~\ref{rangeandnumlayers}(a), we plot the average spatial range of the minimal weight stabilizer operators.
We calculate this quantity by finding the spatial distance between the first and last qubits for each minimal weight stabilizer operator.  We then average over all the minimal weight stabilizer operators for a given state. We further do an ensemble average of this quantity. These minimal weight stabilizer operators are simply the $w$-qubit sets with non-zero correlations found in the first iteration while constructing the entanglement structure diagrams. This $w$ is the erstwhile reported minimal stabilizer weight.  We find that this spatial range grows linearly with system size for unitary volume law states, suggesting the non-local nature of correlations found at the lowest level in their diagrams (see Fig.~\ref{volumelawstruc}(a) for an example). For the measurement-only volume law states, this range does not grow with system size, indicating the presence of local information.  This same locality can also be seen in the
local lowest level clusters %correlations such as 
in the example seen in  Fig.~\ref{volumelawstruc}(b).

Finally, Fig.~\ref{rangeandnumlayers}(b) shows the average number of entanglement structure layers for the volume law states. The number of layers grows with system size more rapidly for measurement-only volume law states than for unitary volume law states. This is reminiscent of the typical case we saw in Fig.~\ref{volumelawstruc} where the measurement-only volume state has a nested structure with a larger number of layers compared to the unitary volume law state.  This is not, however, the best metric for distinguishing the states.

\section{Summary and Outlook}

It is challenging to
understand and visualize  multipartite entanglement, and the literature contains
 a number of  entanglement measures, each of which are useful in appropriate circumstances \cite{MA2024}.  In this paper we 
 introduced a method for
 %constructed 
 %a new way 
  characterizing the multipartite entanglement of stabilizer states.  It allows us to visualize how quantum information and entanglement is internally distributed among the qubits of a many-body state.

The most immediate application of our technique is understanding the volume law states which are produced in random quantum circuits.  None of the traditional entanglement measures are particularly insightful for understanding the properties of these states.  Our approach quantifies the way in which information can be extracted from them when clusters of qubits are interrogated. The size, spatial range and distribution of these clusters helps us identify key differences between volume law states generated by unitary operators versus measurements.   

Rather than using a single number to characterize the entanglement, we produce a diagram, which groups the qubits into a hierarchical arrangement of clusters.  Each cluster is labeled by an integer which specifies the weight of the smallest stabilizer operator connecting the elements contained in it.  From these diagrams we can extract the entanglement depth and bound the entanglement entropy across an arbitrary cut. Our construction provides an alternative way of calculating the exact entanglement depth and all separable partitions of a stabilizer state. It is a computationally efficient procedure for many classes of states.  We can also read off the 'minimal stabilizer weight,' which is the weight of the smallest stabilizer operator. This can also be interpreted as the $k$-uniformity of the state.

The entanglement structure diagrams are trivial for many  classic stabilizer states.   Most states appearing in quantum error correcting codes will have an entanglement structure diagram containing only a single cluster.  The GHZ state also has this structure.  The 1D cluster state with hard wall boundaries, however, has an entanglement structure diagram which reflects the locality of the correlations.

There are several interesting directions that can be explored going forward. In the space of stabilizer states and random quantum circuits, several novel phase transitions and critical states have been found. It would be interesting to resolve structures of critical states and compare different universality classes. We mostly focused on one-dimension (1D) but it is straightforward to extend to higher dimensional states where the spatial structures of correlations can be richer, especially close to criticality \cite{measonly5,measonly6,measonly7,measonly8}. Another avenue which we have not explored is the evolution of the entanglement structure diagram with time. This would give us important insights and visualization of how quantum information gets scrambled under chaotic dynamics from simple initial states. Although our method is best suited for stabilizer states, it would be fruitful to come up with an efficient method to extend the idea of entanglement structure diagrams to non-stablilizer states.  

\section{Acknowledgements}

This material is based upon work supported by the National Science Foundation under Grant No. PHY-2409403. We would like to acknowledge fruitful discussions with Chaoming Jian during this research's formative stages. 

\appendix

\section{Algorithmic details for producing the entanglement structure diagrams}\label{algorithm}

Here we extend the discussion from Sec.~\ref{sec:algebraicstructure}, giving a detailed description of  how we can construct the entanglement structure diagrams for a %arbitrary 
stabilizer state, starting from an arbitrary list of stabilizer generators. 
As described in Sec.~\ref{sec:algebraicstructure}, the key task is to identify $w$-clusters in each iteration until all qubits have been assigned to decoupled large clusters. Each $w$-cluster is constructed by identifying sets where subsets of $w$ elements share non-zero total correlations. 
%We will now describe how total correlations are calculated for stabilizer states and a step-by-step description of the entire algorithm.

In the stabilizer formalism, a pure quantum state of $L$ qubits is described by an $L \times 2L$ binary matrix. The $L$ linearly independent rows encode the Pauli strings which form the stabilizer generators. Each generator can be written in the form, $g = \Pi_{i=1}^{i=L}X_i^{m_i} Z_i^{n_i}$. The $X_i,Z_i$ are Pauli operators for the $i$th qubit and $m_i,n_i$ can be 0 or 1. The quantum state is a simultaneous eigenstate of these $L$ linearly-independent stabilizer generators. This representation is not unique, as the state is unchanged by adding any row to another (mod 2).

In order to form our clusters, we 
must
%first need to be able to %calculate the total correlation $I_A = \sum_i S_i - S_A$.  Here $i$ labels the elements making up the putative cluster, $A$, and $S_i,S_A$ are the entanglement entropies with the rest of the system. 
%calculate the total correlations, we need to first 
calculate the entanglement entropy of groups of $m$ qubits with the rest of the system. %Consider a set of $m$ qubits. 
%This is accomplished by
%We %can 
%find %its 
%it
%the
%entanglement entropy 
%with the rest of the system 
%by %simply
%truncating 
We find this entropy by truncating
the stabilizer matrix, keeping only the columns
%.
%such that we only include columns 
corresponding to these $m$ qubits. 
The resulting reduced stabilizer matrix ($g_m$) is an $L \times 2m$ matrix. The entanglement entropy is given by, $S_m = R - m$ where $R$ is the rank (in modulo 2 arithmetic) of 
%the truncated matrix 
$g_m$~\cite{measonly8}. The value of $R$ can range from $m$ to $2m$. When $R = m$, $S_m = 0$ implying that this $m$-qubit set is disentangled from the rest of the system. Conversely, if $R > m$, this set is entangled with at least some part of the rest of the system.

This procedure can be used to calculate total correlations of any region $A$ consisting of disjoint groups of qubits, where each group is labeled by an integer $i$. The total correlations (multipartite mutual information) are defined as $I_A = \sum_i S_i - S_A$. Here $S_A$ is the entanglement entropy of the entire region $A$ with respect to the rest of the system while $S_i$ is the entanglement entropy of group $i$ %(within $A$) 
with the entire rest of the system. %By calculating the total correlations, we can find $w$-clusters in each iteration using their definition in Sec.~\ref{sec:algebraicstructure}. A $w$-cluster is a region that obeys the following two properties: (a) Any subset of the region containing $(w-1)$ or fewer elements have vanishing total correlations, (b) every element in the region belongs to a subset $C$ containing $w$ elements such that $C$ has non-zero total correlations and $C$ is not disjoint from all such $w$-element subsets within the region.

%We now write down a step-by-step algorithm to find the entanglement structure diagram. During the algorithm we will manipulate two lists $S_1$ and $S_2$...*****  Given an $L$ qubit stabilizer state, we begin by putting each qubit as an 

To calculate our entanglement diagram we begin by placing all of our qubits in a list called \verb+clusters+. Note that in later iterations of the algorithm, each element in \verb+clusters+ can contain multiple qubits. We then follow the steps below:\vspace{0.3cm}

Step 1: Set \verb+n=1+.  This integer will keep track of the cluster weight.\vspace{0.3cm}

Step 2: Calculate entanglement entropy $S$ of each element in \verb+clusters+. %with the rest of the system. 
If $S=0$ for an element, that element is disentangled from the rest of the system. We remove such elements from \verb+clusters+ and include them in the final diagram as isolated elements. The list \verb+clusters+ now has $L_r$ elements remaining. If $L_r = 0$, the process ends. If $L_r \neq 0$, we proceed to the next step.\vspace{0.3cm}

Step 3: Increase \verb+n+ by \verb+1+.\vspace{0.3cm}

Step 4: Form all possible combinations/sets of \verb+n+ elements from \verb+clusters+.  There are $L_r\choose n$ of these. Calculate the total correlations, $I$ of each of these $n$-element sets. If $I = 0$ for all sets, we go back to Step 3 and repeat. If $I \neq 0$ for at least one set, we proceed to Step 5.
We refer to the $I\neq 0$ sets as indivisble.  
\vspace{0.3cm} 

Step 5: 
Remove the elements of the indivisible sets from \verb+clusters+. 
Combine the indivisible sets 
into their disjoint unions, then add these unions back into \verb+clusters+ as new elements.  These are the $n$-clusters corresponding to the  current iteration of the entanglement structure diagram. 
 Return to Step 1.\vspace{0.3cm}
\vspace{0.3cm}
%elements of the $n$ element indivisible sets into disjoint clusters.  Elements are placed into the same cluster if they belong to the same  
%Combine all of the $n$-element sets with non-zero total correlations that have a finite overlap into their union set, and remove their elements from $S_1$. Each union set is an $n$-cluster (containing at least $n$ elements, but possibly more). These union sets are disjoint $n$-clusters. We add these disjoint $n$-clusters as individual elements back to $S_1$.  These new elements are the $n$-clusters seen in a particular iteration of the entanglement structure diagram.\vspace{0.3cm}

%Step 6: 
%We remove all elements from $S_1$ that have some overlap with any element of $S_2$. We then add elements of $S_2$ to $S_1$. The newly added elements from $S_2$ are the $n$-clusters seen in a particular iteration of the entanglement structure diagram. 
%Count the number of remaining elements in the modified list $S_1$ and set it to $L_r$. Move to Step 1.\vspace{0.3cm}

The process ends when the list \verb+clusters+ is empty and at that point, we have the entire entanglement structure diagram. 

\section{Stabilizer group of the 10-qubit state in Fig.~\ref{fig:struct}}\label{app:stabilizers}

Here we write down the stabilizer operators of the 10-qubit stabilizer state example whose entanglement structure was shown in Fig.~\ref{fig:struct} in Sec.~\ref{sec:algebraicstructure}. In some arbitrary gauge, the stabilizer operators of this state are: $S_1 = Y_2Y_3Y_4$, $S_2 = Y_4Y_5X_6$, $S_3 = X_1Y_2Y_5X_6$, $S_4 = X_1Y_2Y_5X_6$, $S_5 = Y_1X_2Z_3Z_5Z_6$, $S_6 = Y_2Y_3X_5Y_6$, $S_7 =X_7X_8X_9X_{10}$, $S_8 = Z_7Z_9$, $S_9 = Z_8Z_9$ and $S_{10} = Z_8Z_{10}$.

\

\section{Scaling of computation time with system size $L$}\label{app:comptime}

The properties of the many-body state determines 
the computational time for the entanglement structure diagram.  In particular, these diagrams are much more costly to calculate for states which contain clusters with large $w$.  The reason for this is two-fold:  First, given $L$ elements, finding indivisible subsets of $w$ elements
requires calculating $L\choose w$ entropies.  Second, each of these entropy calculations requires computing the rank of a $L\times 2w$ matrix, a task whose complexity scales with $w$.

In Sec.~\ref{sec:volume} we encountered two types of volume law states:  those for which the minimal stabilizer weight, $n$, is independent of system size, and those for which $n\propto L$.  For the former, the entanglement structure diagram can be calculated in a time which is polynomial in the system size, scaling as $L^n$.  When $n\propto L$, however, the time scales exponentially with system size as $e^{L\ln L}$.

The algorithm to find the entanglement structure diagram itself reveals the complexity and information scrambling of a quantum state. Although volume law states have diverging entanglement entropy, the level of information scrambling can be markedly different depending on whether it is generated by unitary operators or by projective measurements. We expect that as one increases the proportion of local projective measurements, the computational complexity of our algorithm becomes better.

\bibliography{main}

\end{document}